\documentclass[%
 reprint,
 amsmath,
 amssymb,
 aps,
 nofootinbib,
 prd
]{revtex4-2}

\usepackage{graphicx}
\usepackage{dcolumn}
\usepackage{bm}

\usepackage[T1]{fontenc}
\usepackage[utf8]{inputenc}
\usepackage{amsmath}
\usepackage{mathtools} 
\usepackage{cancel} 
\usepackage{MnSymbol} 
\usepackage{graphicx,adjustbox} 
\usepackage{orcidlink} 
\usepackage{mathrsfs} 
\usepackage{simpler-wick} 
\usepackage{comment}
\usepackage{autobreak}
\usepackage[noend]{algpseudocode}
\usepackage{relsize}
\usepackage{tikz}
\usepackage{float}
\usepackage{stackengine}
\usepackage{booktabs}
\usetikzlibrary{decorations.pathmorphing}
\allowdisplaybreaks
\numberwithin{equation}{section}
\usepackage{nicematrix}
\usepackage{hyperref}
\hypersetup{
	colorlinks=true,
	linktoc=page,
	citecolor=americanrose,
	linkcolor=cadmiumgreen,
	urlcolor=blue} 
\definecolor{gr}{gray}{0.7}

\newcommand{\PM}{\text{PM}}

\usepackage[table]{xcolor}

\definecolor{americanrose}{rgb}{1.0, 0.01, 0.24}
\definecolor{cadmiumgreen}{rgb}{0.0, 0.42, 0.24}

\makeatletter
\newlength{\apb@width}
\newcommand{\autoparbox}[2][c]{\settowidth{\apb@width}{#2}\parbox[#1]{\apb@width}{#2}}
\newcommand{\includegraphicsbox}[2][]{\autoparbox{\includegraphics[#1]{#2}}}

\makeatletter\g@addto@macro\bfseries{\boldmath}\makeatother

\newcommand{\mysumint}{%
  \stackinset{c}{0.5pt}{c}{0pt}{$\scalebox{1.4}[2]{$\int$}$}{$\scalebox{1.3}[1.4]{$\sum$}$}
}

\newcommand{\hdelta}{\hat{\delta}}

\def\e{\mathrm{e}}
\def\D{\mathrm{D}}
\def\d{\mathrm{d}}

\newcommand{\cO}{\mathcal{O}}

\newcommand{\cM}{\mathcal{M}}
\newcommand{\cN}{\mathcal{N}}

\newcommand{\Li}{\operatorname{Li}}

\newcommand{\oldornew}[1]{}

\begin{document}

\title{\boldmath Gravitational Compton scattering at the fifth post-Minkowskian order}

\author{Giacomo Brunello$^{a,b,\,\orcidlink{0009-0004-4788-738X}}$}
\email{giacomo.brunello@sns.it}
\author{Mario Meo$^{a \,\orcidlink{0009-0005-5946-0487}}$}
\email{mario.meo@sns.it}
\author{Sid Smith$^{c,d,e,\, \orcidlink{0009-0007-7799-0136}}$}
\email{sid.smith@ed.ac.uk}

\affiliation{$^a$Scuola Normale Superiore, Piazza dei Cavalieri 7, 56126, Pisa, Italy}
\affiliation{$^b$INFN Sezione di Pisa, Largo
Pontecorvo 3, 56127 Pisa, Italy}
\affiliation{$^c$Dipartimento di Fisica e Astronomia, Università di Padova, Via Marzolo 8, 35131 Padova, Italy}
\affiliation{$^d$INFN, Sezione di Padova,
Via Marzolo 8, I-35131 Padova, Italy.}
\affiliation{$^e$Higgs Centre for Theoretical Physics, University of Edinburgh, James Clerk Maxwell Building, Peter Guthrie Tait Road, Edinburgh, EH9 3FD, United Kingdom}
\date{\today}
\begin{abstract}
Working in the Worldline Quantum Field
Theory framework, we obtain the classical gravitational Compton amplitude through the fifth
post-Minkowskian order, $\mathcal{O}(G^5)$. 
At this order, the point-particle description must be supplemented by the
leading static tidal operators.
After removing lower-order iterations, we extract the
corresponding $N$-matrix element and match it to black-hole perturbation theory.
The matching fixes the static tidal Wilson coefficients to
zero, providing  confirmation of the vanishing static Love numbers of a Schwarzschild black hole.
\end{abstract}
\maketitle
\section{Introduction}
\begin{figure}[t]
  \centering
 \includegraphicsbox[width=0.45\textwidth]{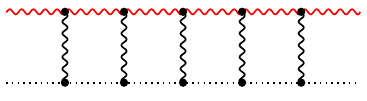}\>
  \caption{
  Four-loop integral topology entering the gravitational Compton amplitude at fifth PM order.
  Black dotted lines denote the compact-object worldline, while black and red wavy lines denote potential and active gravitons respectively.
  }
  \label{fig:four-loop-topology}
\end{figure}
Precision gravitational-wave (GW) physics requires increasingly accurate
theoretical predictions for signals generated by coalescing binary systems.
These signals encode the fundamental properties of the compact objects that
generate them.
Finite-size effects, including tidal deformability, absorption and other
frequency-dependent responses, must therefore be controlled to distinguish
black holes, neutron stars and other compact objects.

Black-hole response theory provides a systematic framework for describing how a
black hole reacts to a passing GW.
The linear response has traditionally been studied using black-hole
perturbation theory (BHPT).
The linear propagation of a GW in a Schwarzschild background is governed by
the Regge--Wheeler and Zerilli
equations~\cite{Regge:1957td,Zerilli:1970se}, while for a Kerr background, one
has to rely on the Teukolsky equation~\cite{Teukolsky:1973ha}.
The phase shifts for these processes can be obtained in partial waves from the
asymptotic solutions of the corresponding differential equations.
Analytic results can be obtained explicitly in a low-frequency expansion, for
example, using the Mano--Suzuki--Takasugi
formalism~\cite{Mano:1996vt,Mano:1996gn}.

Another approach to studying black-hole response relies instead on scattering
amplitudes and worldline quantum field theory (WQFT) methods~\cite{Mogull:2020sak,Jakobsen:2021smu}.
The compact object can be described in an effective field theory (EFT)
framework as a point-like particle moving on a time-like
worldline~\cite{Goldberger:2004jt,Goldberger:2005cd}.

These techniques were originally developed for the study of the gravitational
two-body problem, where they substantially advanced the knowledge of conservative dynamics. 
In the post-Newtonian (PN) expansion, suited to bound and slowly moving systems, multi-loop methods~\cite{Foffa:2016rgu} made it possible to compute the two-body dynamics at 4PN~\cite{Foffa:2019rdf,Foffa:2019yfl} and
5PN~\cite{Foffa:2019hrb,Blumlein:2020pyo,Porto:2024cwd,Porto:2026fsd}, with partial results at
6PN~\cite{Brunello:2025gpf,Almeida:2026clf} and 7PN~\cite{Brunello:2026anu}.
In the relativistic weak field, or post-Minkowskian (PM) expansion, the programme started at
2PM~\cite{Cheung:2018wkq,Bjerrum-Bohr:2018xdl} and moved through
3PM~\cite{Bern:2019nnu,Bern:2019crd,Kalin:2020fhe,Parra-Martinez:2020dzs,Cheung:2020gyp,DiVecchia:2021bdo,Brandhuber:2021eyq}
and
4PM~\cite{Dlapa:2021npj,Bern:2021dqo,Bern:2022jvn,Dlapa:2022lmu,Jakobsen:2023ndj,Jakobsen:2023hig,Damgaard:2023ttc}
up to
5PM~\cite{Bern:2023ccb,Driesse:2024xad,Bern:2024adl,Driesse:2024feo,Bern:2025wyd,Driesse:2026qiz,Dlapa:2026oyq}.
Different formulations can be used to compute classical integrands: via an observable-based scattering amplitudes approach~\cite{Kosower:2018adc}, from
WQFT~\cite{Mogull:2020sak,Jakobsen:2021smu}, or from worldline effective
field theory (WEFT)~\cite{Kalin:2020mvi,Kalin:2022hph}.
The same methods also apply to radiative observables and have been used to compute
gravitational waveforms at
leading~\cite{Jakobsen:2021smu,Mougiakakos:2021ckm,Jakobsen:2021lvp,DeAngelis:2023lvf,Brandhuber:2023hhl,Aoude:2023dui,Falkowski:2024bgb,Brunello:2025cot}
and
next-to-leading~\cite{Brandhuber:2023hhy,Herderschee:2023fxh,Georgoudis:2023lgf,Elkhidir:2023dco,Caron-Huot:2023vxl,Bohnenblust:2023qmy,Brunello:2024ibk,Bohnenblust:2025gir,Brunello:2025eso}
order.

The linear response of the compact object is encoded in the so-called \textit{gravitational Compton}
or \textit{Raman} amplitude.
This is a four-point scattering process in which a graviton scatters off a
compact object and is deflected.
A systematic study of these EFT amplitudes and their direct comparison with
BHPT has recently begun in the PM expansion, by expanding the amplitude in
powers of $\epsilon_{\PM}=2Gm\omega$, where
$\omega$ is the graviton frequency, $m$ is the mass of the compact object, and
$G$ is Newton's constant.
The programme started at 2PM order
($\mathcal{O}(\epsilon_{\PM}^2)$)~\cite{Bjerrum-Bohr:2025bqg}, including spin
effects~\cite{Akpinar:2025huz,Akpinar:2025byi}, and was then extended to 3PM
($\mathcal{O}(\epsilon_{\PM}^3)$)~\cite{Bjerrum-Bohr:2026fhs,Bautista:2026qse,Ivanov:2026icp},
and recently to 4PM
($\mathcal{O}(\epsilon_{\PM}^4)$)~\cite{Brunello:2026rdk,Bautista:2026fcp}.
Once the $N$-matrix is extracted, the 3PM and 4PM results were shown to
reproduce the BHPT phase shift exactly.

The matching of scattering amplitudes to black-hole response has also been
used to study static and dynamical Love
numbers~\cite{Ivanov:2022qqt,Saketh:2023bul,Ivanov:2024sds}.
The separation of the iterations of the long-range gravitational interaction
from the short-distance tidal response is the content of the Born
subtraction~\cite{Correia:2024jgr,Caron-Huot:2025tlq}, which is also the
mechanism underlying the extraction of the $N$-matrix used in this work.

In this work, we compute the gravitational Compton amplitude at
$\mathcal{O}(\epsilon_{\PM}^{5})$, which corresponds to a four-loop calculation.
At this order, static finite-size effects start contributing to the process and are
parametrised in the EFT framework in terms of higher-order operators with
Wilson coefficients describing tidal effects~\cite{Bini:2012gu,Kalin:2020mvi,Jakobsen:2023pvx}.
The leading-order tidal effects describe the gravito-electric and
gravito-magnetic quadrupolar deformation of the compact object.
The integrand is generated recursively~\cite{Berends:1987me,Jakobsen:2023ndj,Jakobsen:2023oow,Bautista:2026qse} within WQFT and reduced, through
integration-by-parts (IBP) identities~\cite{Tkachov:1981wb,Chetyrkin:1981qh,Laporta:2000dsw}, to $30$ master integrals.
The master integrals are evaluated by solving a canonical system of
differential equations (DEs) in the scattering angle~\cite{Kotikov:1990kg,Kotikov:1991pm,Remiddi:1997ny,Gehrmann:1999as,Argeri:2007up,Henn:2013pwa,Argeri:2014qva}, with boundary conditions fixed
in the forward limit using the method of regions~\cite{Beneke:1997zp,Pak:2010pt,Jantzen:2012mw}.
The canonical system contains two elliptic sectors, governed by the same
elliptic curve already found at three loops~\cite{Brunello:2026rdk,Bautista:2026fcp}.
We extract the fifth-order $N$-matrix by subtracting the lower-order
iterations. Unitarity expresses these iterations as rationally weighted
combinations of cuts of the four-loop amplitude.
The subtraction acts directly on the master integrals, so no cut integral needs
to be evaluated explicitly~\cite{Brunello:2026rdk,Bautista:2026fcp}.

Matching its partial-wave decomposition to BHPT fixes the two quadrupolar
Wilson coefficients to zero, in agreement with the vanishing static Love
numbers of a Schwarzschild black
hole~\cite{Fang:2005qq,Damour:2009vw,Binnington:2009bb,Charalambous:2021mea}.
\subsubsection*{Conventions}
We use the mostly-minus metric $\eta^{\mu\nu}=\mathrm{diag}(+,-,-,-)$ and work
in dimensional regularisation with $\D=4-2\epsilon$.
For convenience, we introduce the following shorthand for the $L$-loop integration measure,
\begin{equation}
    \int_{\ell_1\ldots \ell_L}
    \equiv
    \mu^{2\epsilon L}
    \frac{e^{\epsilon\gamma_E L}}{(4\pi)^{\epsilon L}}
    \prod_{r=1}^{L}
    \int\frac{\d^\D\ell_r}{(2\pi)^\D}\,,
\end{equation}
where $\mu$ denotes the renormalisation scale and $\gamma_E$ is the
Euler--Mascheroni constant.

We use bold notation for the $(\D-1)$ spatial components of loop momenta,
\begin{equation}
    \ell^\mu=(\ell^0,\boldsymbol{\ell})\,,
\end{equation}
and a hat to absorb factors of $(2\pi)$,
\begin{equation}
    \hat{\delta}(y)\equiv 2\pi\,\delta(y)\,,
     \qquad
    \hat{\delta}_{+}(\ell^{2})\equiv 2\pi\,\delta(\ell^{2})\,\theta(\ell^{0})\,.
\end{equation}
The gravitational coupling is denoted by $\kappa^2=32\pi G$, with $G$ representing Newton's constant.
\section{Setup of the problem}
\begin{figure}[!ht]
\centering
\includegraphics[width=0.33\textwidth]{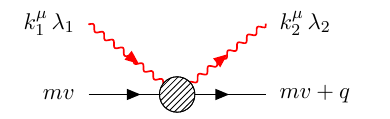}
\vspace{1mm}
\caption{Kinematics for gravitational Compton scattering amplitudes. Incoming and outgoing gravitons carry momenta $k_1^\mu$ and $k_2^\mu$, and helicities $\lambda_1$ and $\lambda_2$, respectively.}
\label{fig:kin-compton-4L}
\end{figure}
The gravitational Compton or Raman scattering
is the elastic scattering of a graviton off a spinless compact object of mass $m$, as illustrated in Fig.~\ref{fig:kin-compton-4L}. 
Defining $\hat T$ as the connected part of the gravitational $S$-matrix,
\begin{equation}
    \hat S = \mathbf{1}+i\,\hat T\,,
    \label{eq:S-T-definition}
\end{equation}
we denote the Compton amplitude as the matrix element of $i\hat T$ between asymptotic one-graviton states,
\begin{equation}
    \label{eq:compton_amplitude}
    \langle k_2,\lambda_2|\,i\hat T\,|k_1,\lambda_1\rangle
    =
    \hat\delta(v\cdot q)\,
    i\mathcal{M}_{\lambda_1\lambda_2}\,.
\end{equation}
Here, $v^\mu$ is the four-velocity of the compact object, with momentum
\begin{equation}
    p^\mu=mv^\mu\,, \qquad v^2=1\,,
\end{equation}
while graviton momenta and helicities are denoted by
\begin{equation}
    (k_i,\lambda_i)\,, \qquad k_i^2=0\,,\qquad i=1,2\,,
\end{equation}
and $q$ is the momentum transferred during the process
\begin{equation}
    q^\mu=k_1^\mu-k_2^\mu\,, \qquad q\cdot v = 0\,.
\end{equation}
This process can be studied in the low-frequency regime,
\begin{equation}
\label{eq:PM_parameter}
    \epsilon_{\PM}=2Gm\omega\,, \qquad \epsilon_{\PM}\ll 1\,,
\end{equation}
where $\omega$ denotes the graviton frequency and $\epsilon_{\PM}$ is the PM
expansion parameter.
In this regime, the radiation wavelength is much larger than the size of the
object, and the process admits an effective field theory (EFT) description. The
compact object is represented by a point particle moving on a timelike
worldline within WQFT~\cite{Mogull:2020sak}.
The amplitude therefore admits the PM expansion
\begin{equation}
    \begin{aligned}
    &i \cM_{\lambda_1\lambda_2}
    =  \sum_{n\ge 1} i \cM^{(n)}_{\lambda_1\lambda_2}\,, \\
    &\cM^{(n)}_{\lambda_1\lambda_2} = \cO\left(\epsilon_{\PM}^n\right)\,.
    \end{aligned}
\end{equation}
In this work, we evaluate the contribution at fifth PM order
$\cM^{(5)}_{\lambda_1\lambda_2}$.
The amplitude contains infrared (IR) divergences, which exponentiate into the Weinberg
phase~\cite{Weinberg:1965nx},
\begin{equation}\label{eq:finiteM-generic}
i\cM=\e^{\,-i\epsilon_{\PM}/\epsilon}\,i\cM_{\mathrm{fin}}\,,
\end{equation}
so that at each PM order the poles are fixed by lower-order amplitudes.
Beyond leading order, the amplitude is not, by itself, the natural object to compare
with general relativity. Not only does it have IR divergences, but it also contains iterations of
the lower PM orders. Both features are removed by passing to the $N$-matrix,
introduced in Sec.~\ref{ssec:N-matrix}.
\subsection{Kinematics of the scattering process}
\label{ssec:kinematics}
In the rest frame of the compact object the external vectors are fixed as
\begin{equation}
    \begin{aligned}
        v &= (1,0,0,0)\,,  \\
        k_{1} &= \omega(1,0,0,1)\,,\\
        k_{2} &= \omega(1,\sin\theta\cos\phi,\sin\theta\sin\phi,\cos\theta)
        \,,
    \end{aligned}
\end{equation}
where $(\theta,\phi)$ specify the direction of the outgoing graviton relative
to the incoming one.
In this frame, the delta function $\hat{\delta}(v\cdot q)$ in
Eq.~\eqref{eq:compton_amplitude} enforces energy conservation, making the
process elastic.
The momentum transfer $q$ can therefore be parametrised in terms of the scattering angle $\theta$,
\begin{equation}
q^2=-4\omega^2\sin^2\frac{\theta}{2}\equiv-4\omega^2x^2\,,
\end{equation}
where we introduced the dimensionless variable
\begin{equation}
    x=\sin\frac{\theta}{2}\,,\qquad 0\le x\le1\,.
\end{equation}
Polarisation tensors are transverse and traceless, and we fix the gauge so that they are orthogonal to the four-velocity $v^\mu$ 
\begin{equation}%
    \label{eq:gauge}
k_{i\mu}\varepsilon^{\mu\nu}_{i,\lambda_i}=0\,,
  \qquad
\eta_{\mu\nu}\varepsilon^{\mu\nu}_{i,\lambda_i}=0\,, \qquad v_\mu\varepsilon^{\mu\nu}_{i,\lambda_i}=0\,.
\end{equation}
They can be written as a double copy of spin-1 polarisation vectors,
\begin{equation}
    \varepsilon^{\mu\nu}_{i,\lambda_i}=\varepsilon^\mu_{i,\lambda_i}\varepsilon^\nu_{i,\lambda_i}\,.
\end{equation}
In the rest frame, a convenient helicity basis with $\lambda_i=\pm1$ is
\begin{equation}
    \begin{aligned}
        \varepsilon_{1,\lambda_{1}} =&\, \frac{1}{\sqrt{2}}(0,1,0,0)+\frac{i\lambda_{1}}{\sqrt{2}}(0,0,1,0)\,,\\
        \varepsilon_{2,\lambda_{2}} =&\, \frac{1}{\sqrt{2}}(0,\cos\theta\cos\phi,\cos\theta\sin\phi,-\sin\theta)\label{eq:polarizations}\\
        &+\frac{i\lambda_{2}}{\sqrt{2}}(0,-\sin\phi,\cos\phi,0)\,.
    \end{aligned}
\end{equation}
\subsection{WQFT framework}
\label{ssec:wqft-frame}
The process is described by the spinless WQFT action~\cite{Mogull:2020sak},
\begin{equation}\label{eq:action-pp}
    S=
    S_{\rm pp}
     +
    S_{\rm EH}
    +
    S_{\rm GF}\,,
\end{equation}
where
\begin{equation}
    \begin{aligned}
        S_{\rm pp}
        &=
        -\frac{m}{2}
        \int \d\tau\,
        g_{\mu\nu}(x(\tau))
        \dot{x}^{\mu}(\tau)
        \dot{x}^{\nu}(\tau)\,, \\
        S_{\rm EH}
        &=
        -\frac{2}{\kappa^2}
        \int \d^\D x\,\sqrt{-g}\,R\,,
        \\ 
        S_{\rm GF}
        &=
        \frac{1}{\kappa^2}
        \int \d^\D x\,\sqrt{-g}\,g^{\mu\nu}\Gamma_\mu \Gamma_\nu\,.
    \end{aligned}
\end{equation}
$S_{\rm pp}$ is the Polyakov action describing the compact-object worldline.
$S_{\rm EH}$ and $S_{\rm GF}$ are, respectively, the Einstein--Hilbert action
and the de Donder gauge-fixing term governing the gravitational dynamics.
Here, $x^\mu(\tau)$ is the worldline parametrised by the proper time $\tau$,
dots denote derivatives with respect to $\tau$, and
$\Gamma_\mu\equiv g^{\rho\sigma}\Gamma_{\rho\sigma\mu}$ is the contracted
Christoffel connection.

The action can be expanded perturbatively around a flat background and the
unperturbed straight-line motion of the object,
\begin{equation}
    g_{\mu\nu}
    =
    \eta_{\mu\nu}
    +
    \kappa \,h_{\mu\nu}\,, \qquad 
    x^\mu(\tau)
    =
    v^\mu\tau
    +
    z^\mu(\tau)\,,
\end{equation}
where $h_{\mu\nu}$ is the graviton field and $z^\mu(\tau)$ is the worldline
fluctuation about the straight trajectory.
Inserting these expansions into the action yields the Feynman rules for
worldline couplings to gravitons and fluctuations $z^\mu(\tau)$, together with
the graviton self-interactions.
Since we compute in-in observables rather than in-out transition amplitudes, the
diagrammatic expansion is organised in the Schwinger--Keldysh
formalism~\cite{Schwinger:1960qe,Keldysh:1964ud}.
Propagators are taken with a retarded $+i\varepsilon$ prescription~\cite{Jakobsen:2022psy,Kalin:2022hph},
\begin{equation}
    \begin{aligned}
        \vcenter{\hbox{\includegraphics[width=0.18\textwidth]{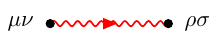}}}
        &\  =\  \frac{i \mathcal{P}_{\mu\nu\rho\sigma}}{(k_0+i\varepsilon)^2-\boldsymbol{k}^2}\,,
        \\ \vcenter{\hbox{\includegraphics[width=0.18\textwidth]{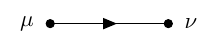}}}  &\ = -\ \frac{i \eta_{\mu\nu}}{m(\omega+i\varepsilon)^2}\,,
    \end{aligned}
\end{equation}
where the $\D$-dimensional graviton polarisation projector is
\begin{equation}
    \mathcal{P}_{\mu\nu\rho\sigma}
    =
    \frac{1}{2}
    \left(
    \eta_{\mu\rho}\eta_{\nu\sigma}
    +\eta_{\mu\sigma}\eta_{\nu\rho}
    -\frac{2}{\D-2}\eta_{\mu\nu}\eta_{\rho\sigma}
    \right)\,.
\end{equation}
This prescription affects only a subset of graviton propagators in the
diagrams.
We refer to these as \textit{active gravitons}.
Potential gravitons carry vanishing energy in the rest frame of the compact
object and are represented by black wavy lines.
Active gravitons carry the external frequency $\omega$, propagate the wave
through the background field and are represented by red wavy lines.

At $L$ loops, the active gravitons form a single chain connecting the incoming to
the outgoing wave and are the only propagators that can go on shell inside the
integration domain.

\subsubsection*{Tidal sector}

The point-particle action in Eq.~\eqref{eq:action-pp} receives corrections from
an infinite tower of effective operators describing finite-size effects of the
compact object at higher orders in perturbation theory.
At the order considered in this work, the leading quadrupolar tidal operators are required.
They can be incorporated into the EFT through the higher-dimensional
action~\cite{Bini:2012gu,Kalin:2020mvi,Mougiakakos:2022sic,Jakobsen:2023pvx}
\begin{equation}\label{eq:tidal-action}
    S_{\rm tidal}
    =
    -\,m\!\int\!\d\tau
    \left[
      \frac{c_{E^2}}{e^3}\,E_{\mu\nu}E^{\mu\nu}
      +\frac{c_{B^2}}{e^3}\,B_{\mu\nu}B^{\mu\nu}
    \right]\,,
\end{equation}
where $e(\tau)$ is an einbein required for reparametrisation invariance.
The operators in Eq.~\eqref{eq:tidal-action} are built from the electric and
magnetic components of the Riemann tensor. On a worldline in vacuum, these
coincide with the corresponding components of the Weyl tensor,
\begin{equation}\label{eq:EB-tensors}
    E_{\mu\nu}=R_{\mu\alpha\nu\beta}\,\dot x^{\alpha}\dot x^{\beta}\,,
    \qquad
    B_{\mu\nu}=R^{\ast}_{\mu\alpha\nu\beta}\,\dot x^{\alpha}\dot x^{\beta}\,,
\end{equation}
with
\begin{equation}
R^{\ast}_{\mu\nu\rho\sigma}=\tfrac12\varepsilon_{\mu\nu\alpha\beta}R^{\alpha\beta}{}_{\rho\sigma}\,.
\end{equation}
The magnetic operator can be continued to $\D$ dimensions by introducing
$B_{\mu\nu\rho}$ through the relations~\cite{Jakobsen:2023ndj}
\begin{equation}
    \begin{aligned}
    B_{\mu \nu \rho} & = \sqrt{\dot{x}^2}\,\dot{x}^\alpha R_{\alpha\mu\nu\rho}\,, \\
    B_{\mu\nu}B^{\mu\nu}& = E_{\mu\nu}E^{\mu\nu}-\frac{1}{2}B_{\mu\nu\rho}B^{\mu\nu\rho}\,.
    \end{aligned}
\end{equation}
The Wilson coefficients $c_{E^2}$ and $c_{B^2}$ encode the leading
quadrupolar tidal response~\cite{Bini:2012gu} and must be fixed by matching to a UV-complete
description of the compact object.
For a Schwarzschild black hole in four dimensions, they are known to
vanish~\cite{Fang:2005qq,Damour:2009vw,Binnington:2009bb,Charalambous:2021mea}.
In four dimensions, the operators $E^2$ and $B^2$ have mass dimension four,
so $c_{E^2}$ and $c_{B^2}$ have mass dimension minus four. Using the
Schwarzschild radius $r_s\sim Gm$ as the characteristic scale, we define the
dimensionless coefficients by
\begin{equation}
    c_{E^2}=(Gm)^4\tilde c_{E^2}\,,
    \qquad
    c_{B^2}=(Gm)^4\tilde c_{B^2}\,.
\end{equation}
The tidal operators first contribute to the Compton amplitude at
$\cO(\epsilon_{\PM}^{5})$, precisely the order considered here.
Expanding $S_{\rm tidal}$ in the graviton field $h_{\mu\nu}$ and the worldline
fluctuation $z^{\mu}$ generates vertices involving two gravitons and additional
worldline fluctuations, with correspondingly higher powers of $\kappa$.

At $\mathcal{O}(G^{5})$, only the two-graviton tidal vertex is
required~\cite{Mougiakakos:2022sic,Ivanov:2026icp}.
Considering both graviton momenta to be outgoing, the Feynman rule is~\cite{Mougiakakos:2022sic}
\begin{equation}\label{eq:tidal-vertex}
\begin{aligned}
\vcenter{\hbox{\includegraphics[width=0.15\textwidth]{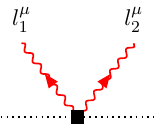}}}
={}&
-\frac{i m\kappa^2}{2}\,
\hdelta\big(v\cdot(\ell_1+\ell_2)\big)
\\[-2pt]
&\hspace{-2.5cm}\times
\left[
c_{E^2}\Pi^E_{\mu\nu,\rho\sigma}(\ell_1,\ell_2)
+c_{B^2}\Pi^B_{\mu\nu,\rho\sigma}(\ell_1,\ell_2)
\right]\,,
\end{aligned}
\end{equation}
where
\begin{equation}
\Pi^X_{\mu\nu,\rho\sigma}(\ell_1,\ell_2)
={}
\mathcal{M}^X_{\mu\nu\mid\alpha\beta}(\ell_1)
{\mathcal{M}^X_{\rho\sigma}}^{\mid\alpha\beta}(\ell_2)\,.
\end{equation}
The first two indices of $\mathcal{M}^X$ are graviton indices, with
$X=E,B$. The electric and magnetic kernels are
\begin{equation}
\begin{aligned}
    \mathcal{M}^E_{\mu\nu\mid\alpha\beta}(\ell)
    ={}&
    v_\mu v_\nu\,\ell_\alpha\ell_\beta
    +(\ell\cdot v)^2\eta_{\mu(\alpha}\eta_{\beta)\nu}
    \\[-2pt]
    &
    -2(\ell\cdot v)\,
    v_{(\mu}\eta_{\nu)(\alpha}\ell_{\beta)}\,,
    \\
    \mathcal{M}^B_{\mu\nu\mid\alpha\beta}(\ell)
    ={}&
    \frac{1}{2}\ell^\gamma v^\delta
    \varepsilon_{\gamma\delta\alpha(\mu}
    \\[-2pt]
    &\times
    \left[
    (\ell\cdot v)\eta_{\nu)\beta}
    -v_{\nu)}\ell_\beta
    \right]
    +(\alpha\leftrightarrow\beta)\,.
\end{aligned}
\label{eq:tidal-electric-magnetic-kernel}
\end{equation}
The magnetic kernel in Eq.~\eqref{eq:tidal-electric-magnetic-kernel} is written in four
dimensions; its continuation to $\D$ dimensions follows from the relation
between $B_{\mu\nu}$ and $B_{\mu\nu\rho}$ given above.
\subsection{\texorpdfstring{$N$}{N}-matrix elements}
\label{ssec:N-matrix}
The amplitude defined in Eq.~\eqref{eq:compton_amplitude} contains iterations of
lower-order processes. These contributions can be removed by introducing the
$N$-matrix operator through the exponential representation of the scattering
matrix~\cite{Damgaard:2021ipf,Bautista:2026qse,Brandhuber:2025igz},
\begin{equation}
\hat S=\e^{\,i\hat N}\,,
\qquad
i\hat N=\log(\mathbf{1}+i\hat T)\,.
\label{eq:T-to-N}
\end{equation}
Expanding both operators in the PM parameter,
\begin{equation}
i\hat T=\sum_{n\geq1}i\hat T^{(n)}\,,
\qquad
i\hat N=\sum_{n\geq1}i\hat N^{(n)}\,,
\end{equation}
the logarithm gives, at a generic order,
\begin{equation}
i\hat N^{(n)}
={}
\sum_{r=1}^{n}\frac{(-1)^{r+1}}{r}
\sum_{\substack{n_1+\cdots+n_r=n\\ n_i\geq1}}\
i\hat T^{(n_1)}\cdots i\hat T^{(n_r)}\,.
\label{eq:N-generic-order}
\end{equation}
The products in Eq.~\eqref{eq:N-generic-order}
are the lower-order Born iterations that must be subtracted from the amplitude~\cite{Correia:2024jgr,Caron-Huot:2025tlq}.
They contain sums over intermediate
on-shell states
\begin{equation}
  X_1\cdot X_2
  =
  \mysumint_{\lambda,\ell}
  \hat\delta_+(\ell^2)\,
  \hat X_1|\ell,\lambda\rangle
  \langle \ell,\lambda|\hat X_2\,.
\end{equation}
We define the matrix element of the $N$-operator as
\begin{equation}
    \langle k_2,\lambda_2|\,i\hat N^{(n)}\,|k_1,\lambda_1\rangle
    ={}
    \hdelta(v\cdot q)\,
    i\mathcal{N}^{(n)}_{\lambda_1\lambda_2}(x,\phi)\,.
    \label{eq:N-matrix-element}
\end{equation}
It is $\cN$, rather than $\cM$, that is the natural object to compare with
BHPT. Unitarity of the $S$-matrix makes $\hat N$ Hermitian. Moreover, the
exponentiated infrared phase in Eq.~\eqref{eq:finiteM-generic} cancels, while the
Born iterations of lower PM orders are subtracted, leaving a finite quantity.

Combining unitarity relations with the Hermiticity of $\hat N$, we can reorganise Eq.~\eqref{eq:N-generic-order} as
 \begin{equation}
 \label{eq:N-generic-unitary}
 \begin{aligned}
 i\hat N^{(n)}
 &=
 \mathrm{Im}\big[i\hat T^{(n)}\big] \\
 &+\sum_{k\geq3}(-1)^{k}\,\frac{k-2}{2\, k}
 \!\!\sum_{\substack{a_1+\cdots+a_k=n\\ a_i\geq1}}\!\!\!\!
 i\hat T^{(a_1)}\!\cdot\ \cdots\ \cdot\, i\hat T^{(a_k)}\,,
 \end{aligned}
 \end{equation}
where the weights depend only on the number of factors $k$. Note that the
terms with $k=2$ cancel identically, so the first subtraction involves three
amplitudes sewn together.
\subsection{Matching to black-hole perturbation theory}
\label{ssec:BHPT-matching}
In BHPT, the same scattering process is described by linear gravitational
perturbations of a Schwarzschild background, decomposed into spherical
harmonics. With the kinematics fixed in Sec.~\ref{ssec:kinematics}, the incoming wave has
azimuthal number equal to $2$. Each multipole splits into odd and even parity sectors,
governed respectively by the Regge--Wheeler ($P=-1$) and Zerilli ($P=+1$) radial
equations~\cite{Regge:1957td,Zerilli:1970se}.
Imposing purely ingoing boundary conditions at the horizon, the radial functions
approach a superposition of ingoing and outgoing waves at large tortoise
coordinate $r_*=r+2Gm\log\!\left(r/(2Gm)-1\right)$,
\begin{equation}\label{eq:BHPT-asymptotics}
    \psi^{P}_{\ell}(r_*)
    \;\xrightarrow[r_*\to\infty]{}\;
    A^{P,\rm in}_{\ell}\,\e^{-i\omega r_*}
    +A^{P,\rm out}_{\ell}\,\e^{+i\omega r_*}\,,
\end{equation}
which defines the partial-wave scattering matrix
\begin{equation}
{}_{-2}S^{P}_{\ell2}
=(-1)^{\ell+1}\,\frac{A^{P,\rm out}_{\ell}}{A^{P,\rm in}_{\ell}}
={}_{-2}\eta_{\ell2}\,
\e^{2i\,{}_{-2}\delta^{P}_{\ell2}}\,,
\label{eq:BHPT-S-matrix}
\end{equation}
where ${}_{-2}\eta_{\ell2}$ is the absorption factor and
${}_{-2}\delta^{P}_{\ell2}$ is the phase shift. The latter admits the
PM expansion
\begin{equation}
{}_{-2}\delta^{P}_{\ell2}
=
\sum_{n\geq1}
\epsilon_{\PM}^{\,n}\,
{}_{-2}\delta^{P,(n)}_{\ell2}\,.
\label{eq:BHPT-phase-expansion}
\end{equation}
%

For a Schwarzschild black
hole, the absorption factor departs from unity only at $\cO(\epsilon_{\PM}^{2\ell+2})$, which is at
$\cO(\epsilon_{\PM}^{6})$ for the dominant quadrupole
$\ell=2$~\cite{Page:1976df}. Consistently, the worldline
action in Eq.~\eqref{eq:action-pp} contains no dissipative
operator~\cite{Goldberger:2005cd}.
Therefore, the absorption factor equals unity at the order considered here.

The WQFT and BHPT results are expressed in different bases.
The $N$-matrix
element depends on the scattering angles and on the graviton helicities, while
the BHPT scattering matrix is diagonal in $\ell$ and parity. 
As shown in
Ref.~\cite{Bautista:2026qse}, they can be compared by projecting the
helicity-preserving and helicity-reversing matrix elements onto spin-weighted
spherical harmonics,
\begin{equation}
\begin{aligned}
A^{(n)}_{\ell}
={}&
\frac{\omega}{4\pi\sqrt{\pi(2\ell+1)}}
\int_{\Omega}{}_{2}Y_{\ell,-2}(\theta,\phi)\,
\overline{\mathcal{N}}^{(n)}_{++}(x,\phi)\,,\\
B^{(n)}_{\ell}
={}&
\frac{\omega}{4\pi\sqrt{\pi(2\ell+1)}}
\int_{\Omega}{}_{2}Y_{\ell,-2}(\pi-\theta,\phi)\,
\overline{\mathcal{N}}^{(n)}_{+-}(x,\phi)\,.
\end{aligned}
\label{eq:N-partial-wave-projection}
\end{equation}
with
\begin{equation}
\int_{\Omega}\equiv\int_{0}^{\pi}\!\d\theta\,\sin\theta\int_{0}^{2\pi}\!\d\phi\,.
\end{equation}
The overline denotes the regularisation of the forward Coulomb singularity~\cite{Bautista:2026qse}. 
The helicity-preserving projection selects the sum of the two parity
phases, while the helicity-reversing projection selects their difference. 
The matching
is then given by~\cite{Bautista:2026qse}
\begin{equation}
\begin{aligned}
A^{(n)}_{\ell}
={}&
\sum_{P=\pm1}{}_{-2}\delta^{P,(n)}_{\ell2}\,,\\
B^{(n)}_{\ell}
={}&
(-1)^{\ell}
\sum_{P=\pm1}P\,{}_{-2}\delta^{P,(n)}_{\ell2}\,.
\end{aligned}
\label{eq:WQFT-BHPT-matching}
\end{equation}
The static quadrupolar response
first contributes at
$\mathcal{O}(\epsilon_{\PM}^{5})$, while absorption and the first dynamical
correction start at $\mathcal{O}(\epsilon_{\PM}^{6})$ and
$\mathcal{O}(\epsilon_{\PM}^{7})$, respectively~\cite{Ivanov:2026icp}.
The matching at $n=5$ is therefore sensitive to the two rescaled static coefficients
$\tilde c_{E^2}$ and $\tilde c_{B^2}$.
\section{Construction of the amplitude}
\label{ssec:constructionoftheamplitude}
We now describe the construction of the fifth post-Minkowskian contribution to
the gravitational Compton amplitude. The calculation closely follows the
fourth-order analysis of Ref.~\cite{Brunello:2026rdk}, so we give only the
details specific to the present case.
\subsection{Integrand generation and tidal contribution}

The recursive construction and the Feynman rules for the point-particle
amplitude follow Ref.~\cite{Brunello:2026rdk}. At 5PM the recursion generates
$248$ WQFT diagrams. The polarisation gauge of Eq.~\eqref{eq:gauge} leaves
$128$ of them non-vanishing.
Tab.~\ref{tab:diagrams} lists the diagram counts up to this order.
\begin{table}[t]
    \centering
    \begin{tabular}{ccc}
        \toprule
        \textbf{~~Loop order $L$~~} & \textbf{~~PM order~~} & \textbf{~~Diagrams~~} \\
        \midrule
        0 & 1PM & $2\,(1)$ \\
        1 & 2PM & $6\,(3)$ \\
        2 & 3PM & $20\,(10)$ \\
        3 & 4PM & $70\,(36)$ \\
        4 & 5PM & $248\,(128)$ \\
        \bottomrule
    \end{tabular}
    \caption{Number of WQFT diagrams produced by the recursion at $L$ loops,
    that is at PM order $n=L+1$. In parentheses we quote how many of them are left
    once the gauge $v_\mu\varepsilon^{\mu\nu}_{i,\lambda_i}=0$ is imposed.}
    \label{tab:diagrams}
\end{table}
The momentum-space integrands were constructed with an in-house
\texttt{FORM}~\cite{Davies:2026cci} implementation that performs the Wick and
index contractions and simplifies the resulting tensors. Throughout, we made use of the on-shell and traceless conditions obeyed by the
external gravitons and the kinematic
identities of Sec.~\ref{ssec:kinematics}. We verified explicitly that the integrand satisfies Ward identities.

The only tidal contribution at $\cO\left(\epsilon_{\PM}^5\right)$ is the
tree-level contact diagram generated by the two-graviton vertex in
Eq.~\eqref{eq:tidal-vertex}, with no internal lines.
After amputating the external legs and contracting with the physical polarisations, this diagram gives
\begin{equation}
\begin{aligned}
&\left.
\langle k_2,\varepsilon_2|\,i\hat T\,|k_1,\varepsilon_1\rangle
\right|_{\rm tidal}
={}
\hdelta(v\cdot q)\,
\varepsilon^{\mu\nu}_{1}
\varepsilon^{*\rho\sigma}_{2}\\
&\hspace{4cm}\times
i\mathcal{M}^{(5)}_{{\rm tidal},\mu\nu,\rho\sigma}\,,
\end{aligned}
\label{eq:tidal-compton-amplitude}
\end{equation}
where
\begin{equation}
    \begin{aligned}
        &i\mathcal{M}^{(5)}_{{\rm tidal},\mu\nu,\rho\sigma}
        ={}
        -\frac{i m\kappa^2}{2}
        \left[
        c_{E^2}\Pi^E_{\mu\nu,\rho\sigma}(-k_1,k_2)
        \right.\\[-2pt]
        &\left.\hspace{3.5cm}
        +c_{B^2}\Pi^B_{\mu\nu,\rho\sigma}(-k_1,k_2)
        \right]\,.
    \end{aligned}
    \label{eq:tidal-M}
\end{equation}
Here, $\Pi^{E}$ and $\Pi^{B}$ are the electric and magnetic structures defined in
Eq.~\eqref{eq:tidal-electric-magnetic-kernel},
evaluated at $\ell_1=-k_1$ and $\ell_2=k_2$, so that the worldline delta
function of Eq.~\eqref{eq:tidal-vertex} reduces to $\hdelta(v\cdot q)$.
The 5PM tidal contribution to the amplitude is reported in the ancillary file \texttt{iM\_tidal\_5pm.m}.

\subsection{Integral reduction}

\begin{figure}[!ht]
\includegraphics[width=0.23\textwidth]{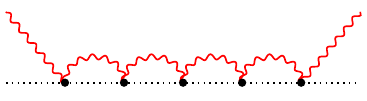}
\includegraphics[width=0.23\textwidth]{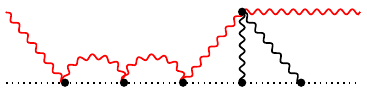}
\vspace{1mm}
\includegraphics[width=0.23\textwidth]{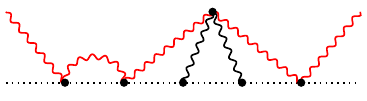}
\includegraphics[width=0.23\textwidth]{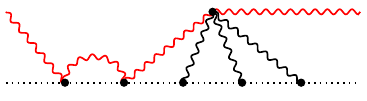}
\vspace{1mm}
\includegraphics[width=0.23\textwidth]{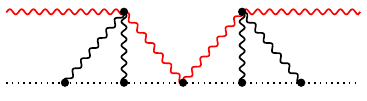}
\includegraphics[width=0.23\textwidth]{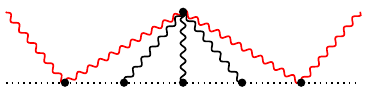}
\vspace{1mm}
\includegraphics[width=0.23\textwidth]{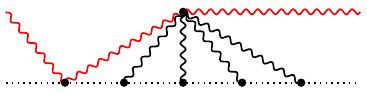}
\includegraphics[width=0.23\textwidth]{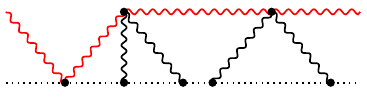}
\vspace{1mm}
\includegraphics[width=0.23\textwidth]{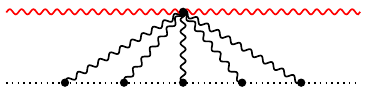}
\includegraphics[width=0.23\textwidth]{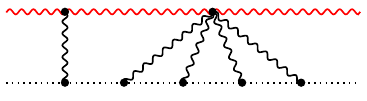}
\vspace{1mm}
\includegraphics[width=0.23\textwidth]{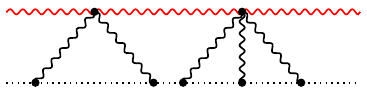}
\includegraphics[width=0.23\textwidth]{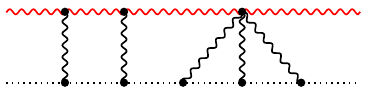}
\vspace{1mm}
\includegraphics[width=0.23\textwidth]{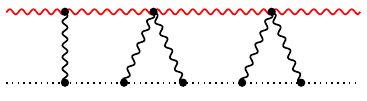}
\includegraphics[width=0.23\textwidth]{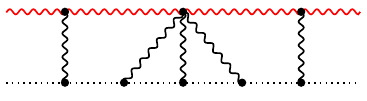}
\vspace{1mm}
\includegraphics[width=0.23\textwidth]{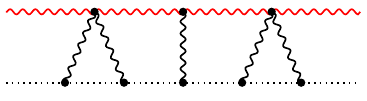}
\includegraphics[width=0.23\textwidth]{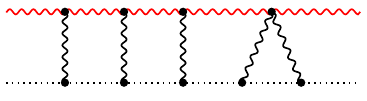}
\vspace{1mm}
\includegraphics[width=0.23\textwidth]{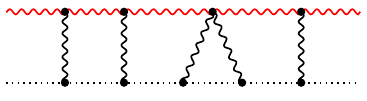}
\includegraphics[width=0.23\textwidth]{figures/master4L_18.pdf}
\vspace{1mm}
   \caption{The 18 sub-topologies $\{\mathcal{G}_i\}_{i=1}^{18}$ of
   the 30 masters of the four-loop reduction. Active graviton propagators are drawn
   in red (see Sec.~\ref{ssec:wqft-frame}) and potential ones in black; worldline
   delta functions appear as black dotted lines.
   }
  \label{fig:four-loop-subgraphs}
\end{figure}

The amplitude contains four-loop tensor integrals up to rank four,
with different momentum routings, worldline delta functions, and polarisation
products $\varepsilon_i\cdot\ell_j$. A Passarino--Veltman (PV) reduction in
\texttt{FORM}~\cite{Anastasiou:2023koq} reduces them to scalar integrals, leaving the
gauge-invariant building blocks
\begin{equation}
    \begin{aligned}
        \mathcal{T}_1
        &\equiv
        \bigl(v\cdot F_1\cdot F_2^*\cdot v\bigr)^2\,,
        \\
        \mathcal{T}_2
        &\equiv
        \bigl(F_1\cdot F_2^*\bigr)
        \bigl(v\cdot F_1\cdot F_2^*\cdot v\bigr)\,,
        \\
        \mathcal{T}_3
        &\equiv
        \bigl(F_1\cdot F_2^*\bigr)^2\,,
    \end{aligned}
\end{equation}
where
$F_i^{\mu\nu}=k_i^\mu\varepsilon_i^\nu-k_i^\nu\varepsilon_i^\mu$. The
worldline delta functions are integrated out, and loop-momentum shifts map all
terms onto a single family of $(\D-1)$-dimensional scalar integrals,
\begin{equation}
    F_{n_{1}\dots n_{18}} = \int_{\boldsymbol{\ell}_{1}\boldsymbol{\ell}_{2}\boldsymbol{\ell}_{3}\boldsymbol{\ell}_{4}}\frac{1}{\prod_{k=1}^{18}D_{k}^{n_{k}}}\,.
\end{equation}
The propagators are
\begin{equation}
    \label{eq:propagators_full}
    \begin{aligned}
    D_{i} &= (\omega+i\varepsilon)^{2}-(\boldsymbol{\ell}_{i}+\boldsymbol{k}_{1})^{2}\,, \quad
    D_{4+i} = -\boldsymbol{\ell}_{i}^{2}\,, \\
    D_{8+i} &= -(\boldsymbol{\ell}_{i}+\boldsymbol{q})^{2}\,, \quad i=1,\dots,4\,,\\
    D_{13} &=-(\boldsymbol{\ell}_{1}-\boldsymbol{\ell}_{2})^{2}\,, \quad D_{14} = -(\boldsymbol{\ell}_{1}-\boldsymbol{\ell}_{3})^{2}\,,\\
    D_{15} &= -(\boldsymbol{\ell}_{1}-\boldsymbol{\ell}_{4})^{2}\,, \quad D_{16} = -(\boldsymbol{\ell}_{2}-\boldsymbol{\ell}_{3})^{2}\,,\\
    D_{17} &= -(\boldsymbol{\ell}_{2}-\boldsymbol{\ell}_{4})^{2}\,, \quad D_{18} = -(\boldsymbol{\ell}_{3}-\boldsymbol{\ell}_{4})^{2}\,,
    \end{aligned}
\end{equation}
where
\begin{equation}
    \boldsymbol{k}_{1}^{2} = \omega^2\,, \quad \boldsymbol{q}^2=4\omega^2 x^2\,, \quad \boldsymbol{k}_{1}\cdot\boldsymbol{q} = 2\omega^{2}x^2\,.
\end{equation}
The resulting family has the kinematics of a three-point process. Only four
propagators depend on the $i\varepsilon$ prescription; these are the active
gravitons. The integrals depend on the kinematic variables $\omega$ and $x$. Since $\omega$ is the
only dimensionful parameter, we set $\omega=1$ during the calculation and
restore it by dimensional analysis.
The scalar integrals are reduced to master integrals through IBP identities
using the program \texttt{PRISM}~\cite{PRISM}. The reduction combines
syzygy-based techniques~\cite{Gluza:2010ws,Wu:2023upw,Wu:2025aeg,Smith:2025xes}
with optimised seeding strategies~\cite{Lange:2025fba,Wu:2023upw,Wu:2025aeg}.
The analytic coefficients are reconstructed over finite fields using
\texttt{FiniteFlow}~\cite{Peraro:2019svx}.
After IBP reduction, the top sector has denominators
$\{D_{1},D_{2},D_{3},D_{4},D_{5},D_{12},D_{13},D_{16},D_{18}\}$, as shown in
Fig.~\ref{fig:four-loop-topology}. The amplitude depends on $30$ master
integrals distributed among the $18$ subsectors in
Fig.~\ref{fig:four-loop-subgraphs}, with multiplicities
\begin{equation}
    \mathcal{G}_{1,\dots,10}:1\,, \quad \mathcal{G}_{11,\ldots,15}:2\,, \quad
    \mathcal{G}_{16,17}:3\,, \quad
    \mathcal{G}_{18}:4\,.
\end{equation}
There are 35 diagrams containing worldline propagators, and they all vanish
after IBP reduction in the TT gauge. This confirms the pattern observed at lower loops in Ref.~\cite{Brunello:2026rdk}.
\subsection{Canonical differential equations}
\label{ssec:canonical differential equation}
\begin{figure}
    \centering
    \includegraphics[width=0.35\textwidth]{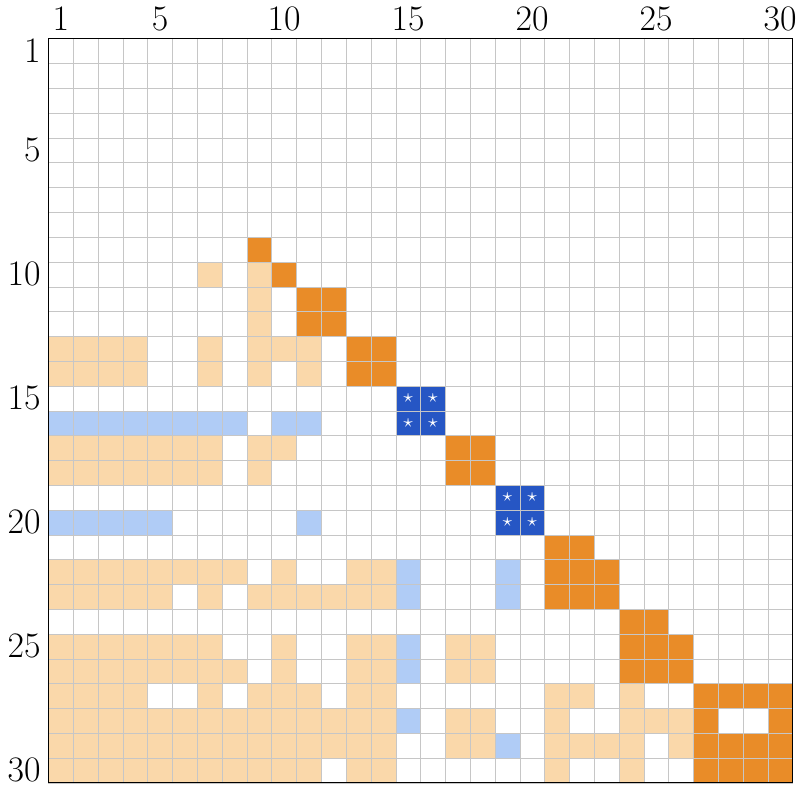}
    \caption{Structure of the $30\times30$ canonical
    connection matrix $\hat{A}(x)$ in Eq.~\eqref{eq:canonical_system}. Rows and
    columns are labelled by the canonical masters $J_1,\dots,J_{30}$. 
    \textbf{Orange:}
    non-elliptic diagonal blocks; 
    \textbf{light orange:} non-elliptic off-diagonal
    couplings; 
    \textbf{blue with a star:} elliptic diagonal sectors $\{J_{15},J_{16}\}$, $\{J_{19},J_{20}\}$;
    \textbf{light blue:} couplings to the
    elliptic sectors. 
    Empty cells vanish.}
    \label{fig:canonical-matrix}
\end{figure}
The master integrals are computed with canonical differential
equations~\cite{Henn:2013pwa} in the single kinematic variable $x$.
In a canonical basis, the dependence on the dimensional regulator $\epsilon$
factorises. Denoting the vector of canonical master integrals by $\mathbf{J}$,
the system reads
\begin{equation}
\label{eq:canonical_system}
\partial_x\mathbf{J}(x,\epsilon)
=
\epsilon \, \hat A(x)\,\mathbf{J}(x,\epsilon)\,.
\end{equation}
We construct the four-loop canonical basis by combining information from leading
singularities~\cite{Flieger:2022xyq,Gorges:2023zgv,Duhr:2025lbz,Forner:2026vby},
Magnus transformations~\cite{Argeri:2014qva}, and the public package
\textsc{Canonica}~\cite{Meyer:2017joq}. The resulting $\epsilon$-factorised
connection $\hat A(x)$ and canonical basis are collected in the ancillary file
\texttt{de\_canonical\_5pm.m}. The two elliptic subsectors
$\mathcal{G}_{13},\mathcal{G}_{15}$ shown in
Fig.~\ref{fig:four-loop-subgraphs} are governed by the same second-order
homogeneous equation as the three-loop problem studied in
Refs.~\cite{Brunello:2026rdk,Bautista:2026fcp},
\begin{equation}
\left(
\partial_x^2-\frac{3x^2-1}{x(1-x^2)}\,\partial_x -\frac{1}{1-x^2}
\right)w_0=0\,.
\end{equation}
A convenient pair of independent solutions is
\begin{equation}
\label{eq:elliptics}
w_0(x)=\frac{2}{\pi}K(x^2)\,, \qquad w_1(x)=K(1-x^2)\,.
\end{equation}
Here, $K$ is the complete elliptic integral of the first kind; both solutions $w_{i}(x)$ are regular for $0< x< 1$.
The structure of the resulting canonical differential equation is shown in Fig.~\ref{fig:canonical-matrix}. This highlights how the elliptic dependence is confined to the masters $\{J_{15},J_{16}\}$ and
$\{J_{19},J_{20}\}$.
The canonical system is solved in terms of iterated integrals over logarithmic
kernels with alphabet $\{x,1+x,1-x\}$~\cite{Chen:1977oja,Remiddi:1999ew},
together with elliptic kernels~\cite{Broedel:2018qkq,Duhr:2019tlz}.
\subsection{Boundary conditions}
\label{ssec:boundary condition}
\begin{figure}[!ht]
    \centering
    \includegraphics[width=0.4\textwidth]{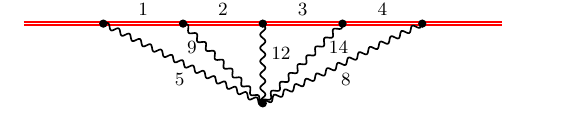}
    \includegraphics[width=0.4\textwidth]{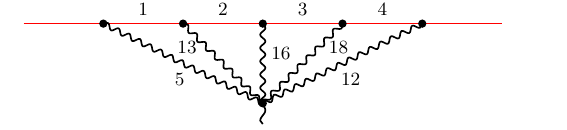}
    \caption{Boundary topologies in the hard (top) and soft (bottom) regions.
    The hard region is a four-loop two-point function with four massive
    propagators, denoted by red double lines. The soft region is a four-loop
    three-point function with four eikonal propagators, denoted by red single
    lines.}
    \label{fig:4loop_regions}
\end{figure}
To solve the canonical system, we determine the boundary vector $\mathbf{J}_b$
in the forward-scattering limit,
\begin{equation}
    x\to0\,.
\end{equation}
At this kinematic point, the differential system is singular, and its
independent asymptotic solutions can be identified through the method of
regions~\cite{Beneke:1997zp,Pak:2010pt,Jantzen:2012mw}. For the four-loop
family, there are two relevant momentum scalings
\begin{equation}
\text{hard scaling:}\quad
\boldsymbol{\ell}_i\sim\omega\,,
\quad
\text{soft scaling:}\quad
\boldsymbol{\ell}_i\sim\omega x\,.
\end{equation}
Rather than expanding every master integral separately, we infer the allowed
scalings directly from the canonical system, following
Refs.~\cite{Mastrolia:2017pfy,Dulat:2014mda,Chestnov:2023kww,Brunello:2025cot}
and the strategy developed for the two- and three-loop analyses in
Refs.~\cite{Bautista:2026qse,Ivanov:2026icp,Brunello:2026rdk}.
The asymptotic
regions follow from the eigenvalue spectrum of the residue matrix at $x=0$.
Bringing this matrix to Jordan form yields
\begin{equation}
M\cdot\epsilon\hat A_0\cdot M^{-1}
=
\mathrm{Diag}
\Bigl[
0^{\times12},
(-8\epsilon)^{\times10},
(2\epsilon)^{\times8}
\Bigr]
\,,
\end{equation}
where $M$ is the Jordan rotation matrix.
The eigenvalues classify the admissible asymptotic behaviours. The sectors
associated with $0$ and $-8\epsilon$ correspond to the hard and soft momentum
regions, where all loop momenta are respectively hard or soft. The $2\epsilon$ branch has no physical region and is removed
by setting its boundary coefficient to zero. 

The structure of the regions agrees with those found at
lower loops~\cite{Bautista:2026qse,Bautista:2026fcp,Brunello:2026rdk,Ivanov:2026icp}. 
The boundary vector reads as
\begin{equation}
  \mathbf{J}_b = \mathbf{J}_0 + x^{-8\epsilon}\,\mathbf{J}_{-8\epsilon}
  \,,
\end{equation}
where $\mathbf{J}_0$ and $\mathbf{J}_{-8\epsilon}$ denote the hard and soft
boundary data. The transformed basis $\mathbf{J}'=M\cdot\mathbf{J}$ diagonalises
the asymptotic behaviour and separates the two contributions. Expanding the
loop integrands in the corresponding regions gives the integral families shown
in Fig.~\ref{fig:4loop_regions}.
\subsubsection{Hard boundary}
\label{ssec:hard boundary}
\begin{figure}[!ht]
\includegraphics[width=0.23\textwidth]{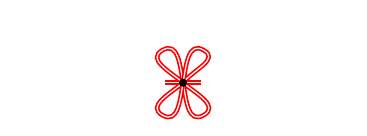}
\includegraphics[width=0.23\textwidth]{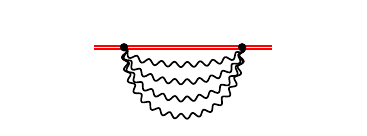}
\vspace{1mm}
\includegraphics[width=0.23\textwidth]{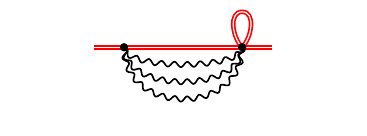}
\includegraphics[width=0.23\textwidth]{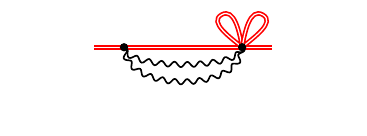}
\vspace{1mm}
\includegraphics[width=0.23\textwidth]{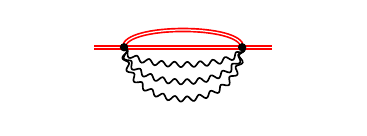}
\includegraphics[width=0.23\textwidth]{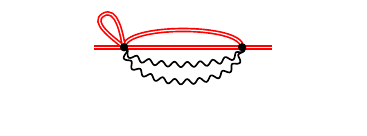}
\vspace{1mm}
\includegraphics[width=0.23\textwidth]{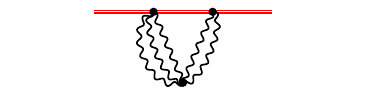}
\includegraphics[width=0.23\textwidth]{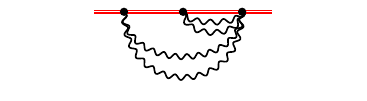}
\vspace{1mm}
\includegraphics[width=0.23\textwidth]{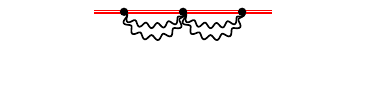}
\includegraphics[width=0.23\textwidth]{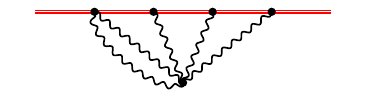}
\vspace{1mm}
\includegraphics[width=0.23\textwidth]{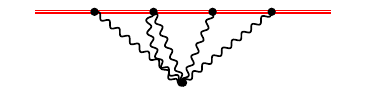}
\includegraphics[width=0.23\textwidth]{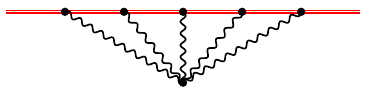}
   \caption{The 12 master integrals in the hard boundary region.}
  \label{fig:hard masters}
\end{figure}
Considering the limit $x\to0$ with all loop momenta hard gives the integral family
\begin{align}
	F_{n_{1}\dots n_{14}}^{\rm h}
	=
	\int_{\boldsymbol{\ell}_1\boldsymbol{\ell}_2\boldsymbol{\ell}_3\boldsymbol{\ell}_4}
  \frac{1}{\prod_{k=1}^{14}(\tilde{D}^{\rm h}_{k})^{n_k}}\,,
\end{align}
where
\begin{equation}
    \label{eq:propagators_hard}
    \begin{aligned}
    \tilde{D}^{\rm h}_{i} &= (\omega+i\varepsilon)^{2}-(\boldsymbol{\ell}_{i}+\boldsymbol{k}_{1})^{2}\,, \\
    \tilde{D}^{\rm h}_{4+i} &= -\boldsymbol{\ell}_{i}^{2}\,, \quad i=1,\dots,4\,, \\
    \tilde{D}^{\rm h}_{9} &=-(\boldsymbol{\ell}_{1}-\boldsymbol{\ell}_{2})^{2}\,, \quad \tilde{D}^{\rm h}_{10} = -(\boldsymbol{\ell}_{1}-\boldsymbol{\ell}_{3})^{2}\,,\\
    \tilde{D}^{\rm h}_{11} &= -(\boldsymbol{\ell}_{1}-\boldsymbol{\ell}_{4})^{2}\,, \quad \tilde{D}^{\rm h}_{12} = -(\boldsymbol{\ell}_{2}-\boldsymbol{\ell}_{3})^{2}\,,\\
    \tilde{D}^{\rm h}_{13} &= -(\boldsymbol{\ell}_{2}-\boldsymbol{\ell}_{4})^{2}\,, \quad \tilde{D}^{\rm h}_{14} = -(\boldsymbol{\ell}_{3}-\boldsymbol{\ell}_{4})^{2}\,.
    \end{aligned}
\end{equation}
The top topology is given by the propagators $\{\tilde{D}^{\rm h}_{1},\tilde{D}^{\rm h}_{2},\tilde{D}^{\rm h}_{3},\tilde{D}^{\rm h}_{4},\tilde{D}^{\rm h}_{5},\tilde{D}^{\rm h}_{8},\tilde{D}^{\rm h}_{9},\tilde{D}^{\rm h}_{12},\tilde{D}^{\rm h}_{14}\}$, shown in Fig.~\ref{fig:4loop_regions}.
The IBP reduction yields the 12 master integrals shown in
Fig.~\ref{fig:hard masters},
\begin{align}
    & F_{11110000000000}^{\rm h}\,, &
    & F_{00011000100101}^{\rm h}\,, &
    & F_{00111000100100}^{\rm h}\,, &\nonumber\\
    & F_{01111000100000}^{\rm h}\,, &
    & F_{10010000100101}^{\rm h}\,, &
    & F_{10110000100100}^{\rm h}\,,
    &\nonumber\\
    & F_{00101001100101}^{\rm h}\,, &
    & F_{01011000100101}^{\rm h}\,, &
    & F_{01101001100001}^{\rm h}\,,
    &\nonumber\\
    & F_{01111001100101}^{\rm h}\,, &
    & F_{10111001100101}^{\rm h}\,, &
    & F_{11111001100101}^{\rm h}\,.
    \label{eq:mis_hard}
\end{align}

\subsubsection{Soft boundary}
\label{ssec:soft boundary}
\begin{figure}[!ht]
\includegraphics[width=0.23\textwidth]{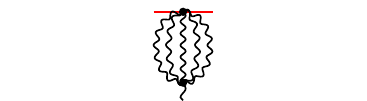}
\includegraphics[width=0.23\textwidth]{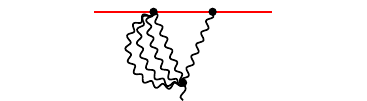}
\vspace{1mm}
\includegraphics[width=0.23\textwidth]{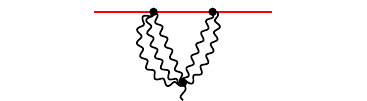}
\includegraphics[width=0.23\textwidth]{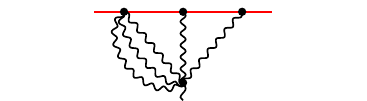}
\vspace{1mm}
\includegraphics[width=0.23\textwidth]{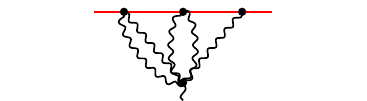}
\includegraphics[width=0.23\textwidth]{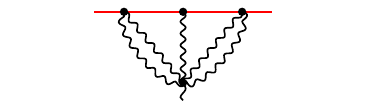}
\vspace{1mm}
\includegraphics[width=0.23\textwidth]{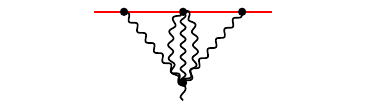}
\includegraphics[width=0.23\textwidth]{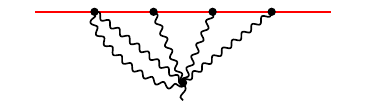}
\vspace{1mm}
\includegraphics[width=0.23\textwidth]{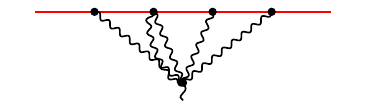}
\includegraphics[width=0.23\textwidth]{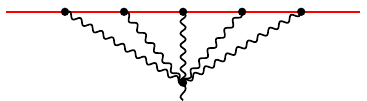}
   \caption{The 10 master integrals in the soft boundary region.}
  \label{fig:soft masters}
\end{figure}
Considering all loop momenta soft instead gives the family
\begin{align}
	F_{n_{1}\dots n_{18}}^{\rm s}=\int_{\boldsymbol{\ell}_1\boldsymbol{\ell}_2\boldsymbol{\ell}_3\boldsymbol{\ell}_4}
  \frac{1}{\prod_{k=1}^{18}(\tilde{D}^{\rm s}_{k})^{n_k}}\,,
\end{align}
where
\begin{equation}
    \label{eq:propagators_soft}
    \begin{aligned}
    \tilde{D}^{\rm s}_{i} &= -2\boldsymbol{\ell}_{i}\cdot\boldsymbol{k}_{1}+i\varepsilon\,, \quad \tilde{D}^{\rm s}_{4+i} = -\boldsymbol{\ell}_{i}^{2}\,, \\
    \tilde{D}^{\rm s}_{8+i} &= -(\boldsymbol{\ell}_{i}+\boldsymbol{q})^{2}\,, \quad i=1,\dots,4\,,\\
    \tilde{D}^{\rm s}_{13} &=-(\boldsymbol{\ell}_{1}-\boldsymbol{\ell}_{2})^{2}\,, \quad \tilde{D}^{\rm s}_{14} = -(\boldsymbol{\ell}_{1}-\boldsymbol{\ell}_{3})^{2}\,,\\
    \tilde{D}^{\rm s}_{15} &= -(\boldsymbol{\ell}_{1}-\boldsymbol{\ell}_{4})^{2}\,, \quad \tilde{D}^{\rm s}_{16} = -(\boldsymbol{\ell}_{2}-\boldsymbol{\ell}_{3})^{2}\,,\\
    \tilde{D}^{\rm s}_{17} &= -(\boldsymbol{\ell}_{2}-\boldsymbol{\ell}_{4})^{2}\,, \quad \tilde{D}^{\rm s}_{18} = -(\boldsymbol{\ell}_{3}-\boldsymbol{\ell}_{4})^{2}\,.
    \end{aligned}
\end{equation}
The top topology is given by the propagators $\{\tilde{D}^{\rm s}_{1},\tilde{D}^{\rm s}_{2},\tilde{D}^{\rm s}_{3},\tilde{D}^{\rm s}_{4},\tilde{D}^{\rm s}_{5},\tilde{D}^{\rm s}_{12},\tilde{D}^{\rm s}_{13},\tilde{D}^{\rm s}_{16},\tilde{D}^{\rm s}_{18}\}$, as shown in Fig.~\ref{fig:4loop_regions}.
The IBP reduction yields the 10 master integrals shown in
Fig.~\ref{fig:soft masters},
\begin{align}
    & F_{000010000001100101}^{\rm s}\,, &
    & F_{000110000001100101}^{\rm s}\,, &\nonumber\\
    & F_{001010000001100101}^{\rm s}\,, &
    & F_{001110000001100101}^{\rm s}\,, &\nonumber\\
    & F_{010110000001100101}^{\rm s}\,, &
    & F_{011010000001100101}^{\rm s}\,, &\nonumber \\
    & F_{100110000001100101}^{\rm s}\,, &
    & F_{011110000001100101}^{\rm s}\,, &\nonumber\\
    & F_{101110000001100101}^{\rm s}\,, &
    & F_{111110000001100101}^{\rm s}\,.
    \label{eq:mis_soft}
\end{align}
Some of these integrals were evaluated analytically in Ref.~\cite{Jinno:2022sbr}.

The boundary vectors $\mathbf{J}_0$ and $\mathbf{J}_{-8\epsilon}$ are independent
of the kinematic variable $x$ and are evaluated numerically with
\textsc{AMFlow}~\cite{Liu:2022chg,Huang:2026rjb} to a precision of up to 300
digits. 
Their analytic form is reconstructed with the PSLQ algorithm using a
suitable basis of transcendental constants implemented in
\textsc{PolyLogTools}~\cite{Duhr:2019tlz}. For this problem, we need transcendental constants up to weight $5$ for the hard region and $4$ for the soft region. 
For the reconstruction, we use transcendental functions:
\begin{equation}
    \left\{\, \pi\,,\ \log2\,, \ \zeta_3\,,  \ \operatorname{Li}_4\!\left(\frac{1}{2}\right) \,, \ \zeta_5  \, \right\} \, .
\end{equation}
The result is provided in the
ancillary file \texttt{retarded\_boundary\_5pm.m}.
\section{Results}

We now present the fifth-order scattering amplitude and the $N$-operator, comparing 
the latter with BHPT.

\subsection{Scattering amplitude}
Expanding Eq.~\eqref{eq:finiteM-generic} at
$\cO\left(\epsilon_{\PM}^5\right)$ gives the IR structure of the four-loop
amplitude,
\begin{align}\label{eq:amplitude_5pm}
    i\mathcal{M}^{(5)}
    =\;&\frac{i\,\epsilon_{\PM}^{4}}{4!\,\epsilon^{4}}\,\mathcal{M}^{(1)}_{0}
    -\frac{\epsilon_{\PM}^{3}}{3!\,\epsilon^{3}}\,\mathcal{M}^{(2)}_{0} \, \nonumber \\
    +&\frac{1}{\epsilon^{2}}\!\left(
        \frac{\epsilon_{\PM}^{3}}{3}\,\mathcal{M}^{(2)}_{1}
        -\frac{i\,\epsilon_{\PM}^{2}}{2}\,\mathcal{M}^{(3)}_{0}
    \right)
    \nonumber\\
    +&\frac{1}{\epsilon}\!\left(
        -\frac{\epsilon_{\PM}^{3}}{3!}\,\mathcal{M}^{(2)}_{2}
        +\frac{i\,\epsilon_{\PM}^{2}}{2}\,\mathcal{M}^{(3)}_{1}+\epsilon_{\PM}\,\mathcal{M}^{(4)}_{0}\right) \nonumber \\
    +&i\mathcal{M}^{(5)}_{0}\,,
\end{align}
where $\cM_k^{(n)}$ denotes the coefficient of $\epsilon^k$ in the $n$PM
contribution.
We checked that the 5PM amplitude satisfies the predicted IR behaviour.
The four-loop scattering amplitude is provided in the ancillary file
\texttt{iM\_5pm.m}.
\subsection{\texorpdfstring{$N$}{N}-operator}
Expanding Eq.~\eqref{eq:N-generic-unitary} at $n=5$ yields 
\begin{align}\label{eq:N5-unitarity}
    i\hat N^{(5)}
    ={}& \mathrm{Im}\big[i\hat T^{(5)}\big]
    \;-\;\frac{1}{6}\!\!\sum_{\substack{a+b+c=5\\ a,b,c\geq1}}\!\!
        i\hat T^{(a)}\!\cdot i\hat T^{(b)}\!\cdot i\hat T^{(c)}
    \nonumber\\[2pt]
    &+\;\frac{1}{4}\!\!\sum_{\substack{a+b+c+d=5\\ a,b,c,d\geq1}}\!\!
        i\hat T^{(a)}\!\cdot i\hat T^{(b)}\!\cdot i\hat T^{(c)}\!\cdot i\hat T^{(d)}
    \nonumber\\[2pt]
    &-\;\frac{3}{10}\,\big(i\hat T^{(1)}\big)^{\cdot\,5}\,,
\end{align}
where there are six terms in the first sum and four in the second.
The iterated contributions in Eq.~\eqref{eq:N5-unitarity} correspond to
unitarity cuts of scattering amplitudes.
Cuts can be applied by replacing the corresponding propagators with their discontinuity~\cite{Anastasiou:2002yz,Anastasiou:2003gr},
\begin{equation}\label{eq:reverse-unitarity}
\frac{1}{p^2}\;\longrightarrow\;
\frac{1}{p^2-i\varepsilon}-\frac{1}{p^2+i\varepsilon}=(2\pi i)\,\delta(p^2)\,.
\end{equation}
Only the active gravitons can go on shell; they correspond to the first four
propagators in Eq.~\eqref{eq:propagators_full}.
The matrix element $\cN^{(5)}$ contains six two-line cuts, four three-line cuts, and a four-line cut,
\begin{align}
\label{eq:N5-cuts}
    i\mathcal{N}^{(5)}
    ={}& \mathrm{Im}\big[i\mathcal{M}^{(5)}\big]
    \;-\;\frac{1}{6}\!\!\sum_{1\leq a<b\leq 4}\!\!\mathrm{Cut}_{(ab)}\big[i\mathcal{M}^{(5)}\big]
    \nonumber\\[2pt]
    &-\;\frac{1}{4}\!\!\sum_{1\leq a<b<c\leq 4}\!\!  \mathrm{Cut}_{(abc)}\big[i\mathcal{M}^{(5)}\big]
    \nonumber \\
    &-\;\frac{3}{10}\,\mathrm{Cut}_{(1234)}\big[i\mathcal{M}^{(5)}\big]\,.
\end{align} 
The extra minus sign on the triple cut contributions is due to differing conventions.
The cuts can be applied directly to the master integrals
\begin{align}\label{eq:cut-sub-masters-5pm}
    \tilde{J}_i
    ={}& \mathrm{Re}\big[J_i\big]
    \;-\;\frac{1}{6}\!\!\sum_{1\leq a<b\leq 4}\!\!
        \mathrm{Cut}_{(ab)}\big[J_i\big]
    \nonumber\\[2pt]
    &
    -\;\frac{1}{4}\!\!\sum_{1\leq a<b<c\leq 4}\!\!
        \mathrm{Cut}_{(abc)}\big[J_i\big]
    \nonumber \\
    &
    -\;\frac{3}{10}\,
        \mathrm{Cut}_{(1234)}\big[J_i\big]\,.
\end{align}
Since the cuts act only on the causality prescription of the propagators, the
canonical system remains unchanged, and the subtraction can be restricted to the
boundary conditions.

The cut-subtracted boundary vector was computed numerically with
\textsc{AMFlow}~\cite{Liu:2022chg,Huang:2026rjb} and reconstructed with
\textsc{PolyLogTools}~\cite{Duhr:2019tlz}, following the same procedure as for
the retarded boundary. It required computing separately the numerical result
for every cut configuration and then combining them as in Eq.~\eqref{eq:cut-sub-masters-5pm}.
It is provided in the ancillary file
\texttt{iterated\_boundary\_5pm.m}.

The point-particle contributions to the 5PM $\mathcal{N}$-matrix elements are
\begin{widetext}
\begin{align}\label{eq:phase_5pm}
\mathcal N^{(5)}_{++}(x,\phi)
={}&\frac{\pi\epsilon_{\PM}^5}{\omega}\,e^{2i\phi}
\Bigg\{
-\frac{x p_1(x)}{11520(1-x^2)^2}
 \big[\mathcal L_2'(x)-(1-x)\mathcal L_2''(x)\big]
-\frac{x^2p_2(x)\log x}{17280(1-x^2)^2}
\notag\\[-2pt]
&\quad
+\frac{x p_3(x)}{276480(1-x^2)}
 \big[\mathcal L_1'(x)+x\mathcal L_1''(x)\big]
+\frac{x p_4(x)}{3(1-x^2)}\mathcal L_1'(x)
+\frac{(1-x)p_5(x)}{15(1-x^2)^2}\mathcal L_2'(x)
\notag\\[-2pt]
&\quad
+\frac{4p_6(x)}{3(1-x^2)^2}\mathcal L_1(x)
+\frac{1}{10x^2}\sum_{\sigma=\pm1}
 \frac{(1-\sigma x)^2}{(1+\sigma x)^2}
 p_7(-\sigma x)\mathcal L_2(\sigma x)
\notag\\[-2pt]
&\quad
+\frac{x p_8(x)w_0(x)\mathcal K'(x)
+\big[xp_8(x)w_0'(x)+p_9(x)w_0(x)\big]\mathcal K(x)}{768(1-x^2)}
-\frac{p_{\pi}(x)}{207360(1-x^2)^2}
\Bigg\}\,,
\notag\\[4pt]
\mathcal N^{(5)}_{+-}(x,\phi)
={}&\frac{\pi\epsilon_{\PM}^5}{\omega x^2}\,e^{2i\phi}
\Bigg\{
\frac{p_{10}(x)}{1620}
-\frac{p_{11}(x)\log x}{135}
+\frac{p_{12}(x)}{45x}
 \big[\mathcal L_2'(x)-(1-x)\mathcal L_2''(x)\big]
\notag\\[-2pt]
&\quad
+\frac{16(1-x)p_{13}(x)}{15}\mathcal L_2'(x)
+\frac{8p_{14}(x)}{5x^2}
 \big[\mathcal L_2(x)+\mathcal L_2(-x)\big]
\Bigg\}
\,.
\end{align}
\end{widetext}
where we have defined 15 polynomials
\begin{widetext}
\begin{minipage}[t]{0.48\textwidth}
    \begin{align}
        p_1(x) = {}&3915542+7778692x^2-9092631x^4
        \nonumber
        \\
        &-2596878x^6-4725x^8
        \nonumber\,,
        \\
        p_3(x) = {}&556170+2775515x^2+779892x^4+1575x^6
        \nonumber\,,
        \\
        p_5(x) = {}&-135+246x^2-231x^4+176x^6
        \nonumber\,,
        \\
        p_7(x)={}&16+11x-44x^2+11x^3+16x^4
        \nonumber\,,
        \\
        p_9(x)={}&29976+120355x^2+25485x^4
        \nonumber\,,
        \\
        p_{11}(x)={}&270060-166200x^2+8093x^4
        \nonumber\,,
        \\
        p_{13}(x)={}&1140-528x^2+11x^4
        \nonumber\,,
    \end{align}
\end{minipage}
\hfill
\vspace{-6pt}
\begin{minipage}[t]{0.48\textwidth}
    \begin{align}
        p_2(x) = {}&4205883+8476650x^2+904459x^4
        \nonumber\,,
        \\
        \nonumber\\
        p_4(x)={}&7+11x^2
        \nonumber\,,
        \\
        p_6(x)={}&1+6x^2+2x^4
        \nonumber\,,
        \\
        p_8(x)={}&88546+227476x^2+35610x^4
        \label{eq:p_polynomials}\,,
        \\
        p_{10}(x)={}&1620360-1279170x^2+96979x^4
        \nonumber\,,
        \\
        p_{12}(x)={}&-90020+114090x^2-33792x^4+1017x^6
        \nonumber\,,
        \\
        p_{14}(x)={}&-760+732x^2-120x^4+x^6
        \nonumber\,,
    \end{align}
\end{minipage}
\begin{align}
    p_{\pi}(x) = {}&15577856+2167296\pi^2+82944\pi^4+(40280868+7259490\pi^2+635904\pi^4)x^2
    \nonumber\\
    &-(45773352+6944145\pi^2-184320\pi^4)x^4\quad-(10085372+2477916\pi^2)x^6-4725\pi^2x^8
    \,.\nonumber
\end{align}
\end{widetext}
The result is finite and expressed in terms of two weight four polylogarithmic functions and their derivatives
\begin{equation}
    \begin{aligned}
         \mathcal L_1(x)={}
        &
        24\big(G_{0,-1,1,0}(x)+G_{0,1,-1,0}(x)\big)
        \\[-2pt]
        &{}-12\log x\,\Li_3(x^2)+18\Li_4(x^2)
        \\[-2pt]
        &{}-\pi^2\Li_2(x^2)
        \, ,
        \\[2pt]
        \mathcal L_2(x)={}&
        G_{1,0,-1,0}(x)+G_{1,0,1,0}(x)
        \\[-2pt]
        &{}-\frac12\zeta_3\log(1-x)
        \, .
    \end{aligned}
\end{equation}
The elliptic dependence is contained in the function
\begin{equation}
    \begin{aligned}
        \mathcal K(x)={}&
        \frac{w_1(x)}{w_0(x)}
        \int_0^x\!\mathrm d x'\,\Omega(x')w_0(x')
        \\[-2pt]
        &{}-\int_0^x\!\mathrm d x'\,\Omega(x')w_1(x')
        +\frac{\pi^4}{16}
        \,,
    \end{aligned}
\end{equation}
where
\begin{equation}
\begin{aligned}
\Omega(x')={}&
\Li_2(x')-\Li_2(-x')
+\log x'\log\!\frac{1-x'}{1+x'}
\\[-2pt]
&{}+\frac{\Li_2(x'^2)+2\log x'\log(1-x'^2)}{x'}
\,,
\end{aligned}
\end{equation}
and $w_i(x)$ are defined in Eq.~\eqref{eq:elliptics}.
The functions $G_{a_1,\ldots,a_n}$ are Goncharov
polylogarithms~\cite{Goncharov:1998kja}, defined recursively by
\begin{equation}
\begin{aligned}
G_{a_1,\ldots,a_n}(x)
&=
\int_0^x\frac{\d x'}{x'-a_1}\,
G_{a_2,\ldots,a_n}(x')
\,,
\\
G_{\emptyset}(x)&=1\,,
\end{aligned}
\end{equation}
The letters belong to the alphabet $a_i\in\{-1,0,1\}$, and the integration
contour is the real segment $0\leq x'\leq x<1$.

These point-particle matrix elements are collected in the ancillary file
\texttt{N\_matrix\_5pm.m}.

Since the tidal interaction first enters at 5PM, none of the lower-order Born
iterations contains a tidal insertion. Its contribution to the $N$-matrix
therefore coincides with the corresponding amplitude and reads
\begin{equation}\label{eq:phase_5pm_tidal}
\begin{aligned}
\left.\mathcal N^{(5)}_{++}(x,\phi)\right|_{\rm tidal}
={}&
-\frac{\pi\epsilon_{\PM}^5}{2\omega}\,e^{2i\phi}
\big(\tilde c_{E^2}+\tilde c_{B^2}\big)(1-x^2)^2
\,,
\\[2pt]
\left.\mathcal N^{(5)}_{+-}(x,\phi)\right|_{\rm tidal}
={}&
-\frac{\pi\epsilon_{\PM}^5}{2\omega}\,e^{2i\phi}
\big(\tilde c_{E^2}-\tilde c_{B^2}\big)x^4
\,.
\end{aligned}
\end{equation}
The tidal $N$-matrix elements are collected in the ancillary file
\texttt{N\_matrix\_tidal\_5pm.m}.
\subsection{Matching to BHPT and tidal contribution}
\label{ssec:matching-tidal}
The full 5PM matrix elements are obtained by adding the point-particle
contributions in Eq.~\eqref{eq:phase_5pm} and the tidal contributions in
Eq.~\eqref{eq:phase_5pm_tidal}. 
They can be matched to BHPT following the
procedure outlined in Sec.~\ref{ssec:BHPT-matching}.

We project $\cN^{(5)}$ onto spin-weighted spherical harmonics using
Eq.~\eqref{eq:N-partial-wave-projection} and compare the two parity combinations
with the BHPT phase shifts through Eq.~\eqref{eq:WQFT-BHPT-matching}.
The BHPT phase shifts are obtained in a low-frequency expansion using publicly
available codes~\cite{BlackHolePerturbationToolkit,Markovic:2025kvr}.
We evaluate the projections for $\ell=2,\ldots,20$.
The angular dependence of
the tidal contact terms is entirely carried by the $\ell=2$ spin-weighted
spherical harmonics, and they do not contribute to higher multipoles.
The $\ell=2$ projections read
\begin{equation}
\begin{aligned}
   \left. A_2^{(5)}\right|_{\rm tidal} 
    & = -\frac{1}{20}\left(\tilde c_{E^2}+ \tilde c_{B^2}\right)
    \, , \\
   \left. B_2^{(5)}\right|_{\rm tidal}  
    & =-\frac{1}{20}\left(\tilde c_{E^2}- \tilde c_{B^2}\right)
\end{aligned}    
\end{equation}
For $\ell\ge 3$, the matching therefore probes the point-particle dynamics
alone, and Eq.~\eqref{eq:WQFT-BHPT-matching} is satisfied in both helicity sectors.

At $\ell=2$, the two helicity projections fix the sum and difference of the two
rescaled Wilson coefficients, giving
\begin{align}\label{eq:love-zero}
\tilde c_{E^2}=\tilde c_{B^2}=0\,,
\end{align}
in agreement with the vanishing static Love numbers of a Schwarzschild black
hole~\cite{Fang:2005qq,Damour:2009vw,Binnington:2009bb,Charalambous:2021mea}.
This provides an on-shell, gauge-invariant confirmation of this result at
$\cO(\epsilon_{\PM}^{5})$.
\section{Outlook}
This work provides the classical gravitational Compton amplitude through
$\cO(G^5)$, obtained within WQFT.
The four-loop WQFT integrand is constructed recursively, and it contains 248 diagrams.
Using IBP reduction, the amplitude is decomposed in terms of $30$ master integrals, which
we evaluated using canonical differential equations.
After subtracting lower-order Born iterations, we extracted the corresponding
$N$-matrix element for the point-particle contribution.
Static finite-size effects first enter at this order through the leading quadrupolar tidal operators.
We computed the corresponding tidal contribution to the $N$-matrix element.

We found agreement with BHPT for the total $N$-matrix element. The tidal part contributes only to the $\ell=2$ partial wave,
which fixes both static tidal Wilson coefficients to zero.
This provides confirmation of the
vanishing static Love numbers of a Schwarzschild black hole.

Several directions follow naturally.
The same integral family appears for spinning binary systems. The integrals computed in this work can therefore be reused in that case.
A direct matching in that setting would probe Kerr Love numbers beyond the orders currently known.

Extending this setup to higher orders allows us to capture absorption effects and the
leading dynamical response, which enter for the first time at $\cO(\epsilon_{\PM}^{6})$ and
$\cO(\epsilon_{\PM}^{7})$, respectively~\cite{Ivanov:2026icp}.

Finally, the appearance of two elliptic sectors at four loops, governed by the
same curve already found at three loops, suggests that the function space of
gravitational-wave scattering is more constrained than the loop order alone
would indicate. 
Understanding this structure will be valuable for extending the calculation to
higher orders.
\begin{acknowledgments}
We wish to thank Fabian Bautista, Stefano De Angelis, Mathias Driesse, Kays Haddad, Gustav Jakobsen, and Pierpaolo Mastrolia for useful discussions.
We also thank Xiao Liu and Yan-Qing Ma for their invaluable help with \textsc{AMFlow}.
The computations for this work used resources provided by the CloudVeneto
initiative at the University of Padova and INFN, the Magellano HPC cluster of
INFN Pisa, and the Hoffman2 Cluster, operated by the UCLA Office of Advanced
Research Computing's Research Technology Group.
G.B.'s research is supported by the Italian MIUR under contract 20223ANFHR (PRIN2022), by the ERC (NOTIMEFORCOSMO, 101126304),
and by the INFN initiatives \textit{Amplitudes} and \textit{TPPC}.
S.S.'s research is partially supported by the INFN initiative
\textit{Amplitudes}.
\end{acknowledgments}
\bibliographystyle{apsrev4-1}
\bibliography{binary.bib}

\onecolumngrid
\newpage

\end{document}